\documentstyle[twoside,fleqn,espcrc2]{article}
\begin{document}
\newcommand{\ttbs}{\char'134}
\newcommand{\ubar}{\bar{U}}
\newcommand{\cm}{Commun.\ Math.\ Phys.~}
\newcommand{\pr}{Phys.\ Rev.\ D~}
\newcommand{\pl}{Phys.\ Lett.\ B~}
\newcommand{\ibar}{\bar{\imath}}
\newcommand{\jbar}{\bar{\jmath}}
\newcommand{\taubar}{\bar{\tau}}
\newcommand{\I}{{\cal I}}
\newcommand{\F}{{\cal F}}
\newcommand{\R}{{\cal R}}
\newcommand{\A}{{\cal A}}
\newcommand{\M}{{\cal M}}
\newcommand{\K}{{\cal K}}
\newcommand{\T}{{\cal T}}
\newcommand{\np}{Nucl.\ Phys.\ B~}
\newcommand{\AmS}{{\protect\the\textfont2
  A\kern-.1667em\lower.5ex\hbox{M}\kern-.125emS}}

\newcommand{\be}{\begin{equation}}
\newcommand{\ee}{\end{equation}}
\newcommand{\ba}{\begin{eqnarray}}
\newcommand{\ea}{\end{eqnarray}}
\begin{titlepage}
\begin{flushright}
\large
CERN--TH/97-136\\[2mm]
CPTH--5370.0697\\[2mm]
hep-th/9706211
\end{flushright}\vskip 1cm
\begin{center}\Large\bf Lectures on Heterotic--Type I 
Duality$^*$\\[1cm]
\Large I.Antoniadis,$^{a,b}$ H. Partouche,$^b$ T.R. Taylor$^c$\\[1cm]
\large\sl
$^a$TH-Division, CERN, CH-1211 Geneva 23, Switzerland\\[1mm] \large\sl
$^b$Centre de Physique Th{\'e}orique, Ecole Polytechnique, F-91128 
Palaiseau cedex, France\\[1mm] \large\sl
$^c$Department of Physics, Northeastern University, Boston, MA 02115,
U.S.A.\\[1cm]
\bf Abstract
\end{center}
{\large We present a review of heterotic--type I string duality. In particular, 
we discuss the effective field theory of six- and four-dimensional
compactifications with $N>1$ supersymmetries. We then describe various duality
tests by comparing gauge couplings, $N=2$ prepotentials, as well as 
higher-derivative F-terms.\\[1cm]
\it
Based on invited lectures delivered at:
33rd Karpacz Winter School of Theoretical Physics
``Duality, Strings and Fields,''
Przesieka, Poland, 13 - 22 February 1997; 
Trieste Conference on Duality Symmetries in String Theory,
Trieste, Italy, 1 - 4 April 1997;
Carg\`ese Summer School ``Strings, Branes and Dualities,''
Carg\`ese, France, 26 May - 14 June 1997}.\\[3cm]
\hrule\vskip 2mm\noindent
$^{*}$Research supported in part by the 
National Science Foundation under grant PHY--96--02074, and in part by the EEC 
under the TMR contract ERBFMRX-CT96-0090.
\begin{flushleft}
\large
CERN--TH/97-136\\[3mm]
June 1997
\end{flushleft}
\end{titlepage}   
%
%f
% add words to TeX's hyphenation exception list
\hyphenation{author another created financial paper re-commend-ed}
% declarations for front matter
\title{Lectures on Heterotic--Type I 
Duality\thanks{Research supported in part by the 
National Science Foundation under grant PHY--96--02074, and in part by the EEC 
under the TMR contract ERBFMRX-CT96-0090.}}
\author{I. Antoniadis\address{TH-Division, CERN, CH-1211 Geneva 23, 
        Switzerland}$^{\mbox{\scriptsize ,b}}$, 
        H. Partouche\address{Centre de Physique
        Th{\'e}orique, Ecole Polytechnique, F-91128 Palaiseau cedex,
        France} and T.R. Taylor\address{Department of Physics, 
        Northeastern University, Boston, MA 02115, U.S.A.}}
%\begin{document}
\begin{abstract}
We present a review of heterotic--type I string duality. In particular, 
we discuss the effective field theory of six- and four-dimensional
compactifications with $N>1$ supersymmetries. We then describe various duality
tests by comparing gauge couplings, $N=2$ prepotentials, as well as 
higher-derivative F-terms.

%\today
\end{abstract}

\maketitle
\section{INTRODUCTION}

In the last couple of years, there has been remarkable progress in 
understanding non-perturbative aspects of 
string theories. The new ingredient is the
discovery of various duality symmetries; these, in particular, invert the
coupling constant and exchange perturbative states with non-perturbative
solitonic excitations. Since inversion of the coupling involves the Planck 
constant $\hbar$, duality symmetries can be realized only at the quantum level.
Once the dual models are identified,
non-perturbative dynamics of one model
can be studied by means of the standard perturbative techniques
in the other model.

Dualities are known under the names of 
S, T and U, depending on whether they invert 
the string coupling in the target space-time, or the coupling of the 
underlying two-dimensional field theory of the string world-sheet, 
or exchange these two, respectively \cite{Sch}. 
One of the main consequences of string 
dualities is that all known, apparently distinct, superstring theories
correspond to different perturbative expansions of the same
underlying theory, the so-called M-theory \cite{Wvar}, whose low-energy limit 
is eleven-dimensional supergravity. In fact, there are 
five consistent superstring
theories in ten dimensions: two type II theories of closed strings, IIA and IIB,
which have two supersymmetries in ten dimensions, of opposite and 
same chirality,
respectively; two heterotic closed string theories with $N=1$ supersymmetry 
and $SO(32)$ or $E_8\times E_8$ gauge groups; and the type I theory of open and
closed strings with $N=1$ supersymmetry and $SO(32)$ gauge group \cite{gsw}. 

It was already
known for quite some time that the two type II theories, as well as the two
heterotic ones, are related by T-duality, which is an exact perturbative
symmetry after compactifying to lower dimensions. For instance, 
upon compactification to nine dimensions on a circle of radius $R$,
type IIA and type IIB theories become equivalent by inverting $R$:
\be
R\stackrel{T}{\longleftrightarrow}{\alpha'\over R}\, ,
\label{R1R}
\ee
where $\alpha'$ is the Regge slope. In four dimensions, the two theories are 
also identified by mirror symmetry with respect to the internal six-dimensional
Calabi--Yau manifold. T-duality can then be used to compute world-sheet 
instanton corrections by using algebraic geometry in the mirror (dual) theory
\cite{COFKM}.
Similarly, the two heterotic theories are equivalent but T-duality is more
complicated due to the presence of Wilson lines. 

It has recently been realized that, in ten dimensions, there are two fundamental
non-perturbative S-dualities. On the one hand, the heterotic and type I
$SO(32)$ theories are equivalent by inverting the string coupling 
\cite{Wvar,PW}:
\be
\lambda\stackrel{S}{\longleftrightarrow}{1\over \lambda}\, .
\ee
We recall that the string coupling is
given by the vacuum expectation value (vev) of the 
dilaton field:
$\lambda=\langle e^{\phi}\rangle$.
On the other hand, type IIB is self-dual under an $SL(2;{\bf Z})$ S-duality 
\cite{ht} acting on the complex field
$\chi+i\lambda$, where $\chi$ is the Ramond--Ramond (RR) scalar. 
In this way, the
two $N=2$ and three $N=1$ theories are separately connected. Their mutual
connection is believed to arise from U-duality via M-theory \cite{Wvar,HW}. 
In fact, the coupling constants of the type
IIA and heterotic $E_8\times E_8$ strings can be identified with the
radius of the eleventh dimension of the M-theory compactified on $S^1$ and
$S^1/{\bf Z}_2$, respectively:
\be
\lambda = (R_{11}/l_P)^{3/2}\, ,
\ee
where $l_P$ is the Planck length in eleven dimensions. 

The purpose of these notes is to describe the main features, consequences and
tests of heterotic--type I string duality. In Section 2, we give a brief
introduction to type I strings, D-branes and orientifolds.
In Section 3, we derive the main
duality conjectures in dimensions 10, 6 and 4 by arguments based on the
effective field theory. In Section 4, we discuss duality in six dimensions
with $N=1$ supersymmetry, while in Section 5, we study four-dimensional
compactifications with $N=2$ supersymmetry. In Section 6, we present general
results for the one-loop corrections to the $N=2$ prepotential and vector moduli
metrics. We then perform various duality tests in the Higgs (Section 7.1) and
Coulomb (Section 7.2) phase of $(T^4/{\bf Z}_2)\times T^2$ compactifications. 
Finally, in Section 8, we compare the infinite series of couplings corresponding
to higher-derivative F-terms.
 
\section{TYPE I STRINGS AND D-BRANES}

Type I strings arise as world-sheet orbifolds (orientifolds)
of the world-sheet left--right-symmetric IIB theory by
modding it out with respect to the
involution $\Omega$ that exchanges left- and right-movers \cite{bs,hor}. The
projection in the ``untwisted" closed string sector introduces unoriented closed
strings and has the effect of symmetrizing the 
Neveu--Schwarz--Neveu--Schwarz (NSNS)
sector and antisymmetrizing the RR sector. As a result, the
number of supersymmetries is reduced by half. In ten dimensions, the invariant
(bosonic) massless states are the dilaton and graviton of the NSNS sector and
the 2-index antisymmetric tensor of the RR sector.

Open strings appear as type IIB strings that close up
to an $\Omega$ reflection, in a way similar to ``twisted''
states in orbifold constructions:
\begin{equation}
\Omega X(\sigma,\tau)\equiv X(-\sigma,\tau)=X(\sigma,\tau)\, .
\label{twist}
\end{equation}
Indeed, the mode expansion of an $\Omega$-invariant 
free string satisfying Eq.\  (\ref{twist}) is
\ba
X(\sigma,\tau)&=& x+ 2\alpha'p\tau\label{mode}\\
&+&\!\! {i}\sqrt{2\alpha'}
\sum_{m\neq 0}\frac{\alpha_m}{m}
\cos(m\sigma)e^{-im\tau},\nonumber
\ea
which is the usual open string with Neumann boundary conditions
$\partial_\sigma X|_{\sigma=0,\pi}=0$.
Open strings carry at their ends Chan--Paton charges that give rise to a gauge
group. From the orientifold point of view, they 
play the role of fixed points. Their
multiplicity is constrained by the tadpole cancellation which replaces
(oriented) closed string modular invariance. In ten dimensions, one finds 32
real charges leading to an $SO(32)$ gauge group. 

The modern approach to type I theory emphasizes the role
of p-brane solitons of the corresponding type II string theory.
In particular, the multiplicity of (complex) charges corresponds to the
number of the so-called Dp-branes where open strings can end.
Dp-branes provide sources for the RR type II (p+1)-potentials \cite{P}.
An open string with one end (say at $\sigma=0$) on
a Dp-brane has Neumann boundary conditions for $X^\mu$ 
with $\mu = 0,1,\dots ,$p and Dirichlet boundary conditions along the
perpendicular directions, $\partial_\tau X^\mu|_{\sigma =0}=0$ 
for $\mu={\rm p}+1,\dots,9$. These conditions fix the position
of the string end in the space orthogonal to the p-brane, allowing it to move
freely only in the directions parallel to the p-brane.
According to D-brane terminology, strings with one end on a Dp-brane and
the other one on a Dq-brane are called pq strings.

Type IIB has even-form RR
potentials that can couple to Dp-branes with p odd. The
$\Omega$ projection leaves invariant 2 mod 4 forms coupled to D1-, D5-
and D9-branes only. In ten dimensions, Lorentz invariance 
allows only D9-branes with multiplicity 16 fixed by the tadpole cancellation.
However after compactification to six or four dimensions, D5-branes can also
appear.

Following the discussion above, the 1-loop partition
function receives contributions from four genus-one surfaces: the torus 
($\T$), the Klein bottle ($\K$), the annulus ($\A$) and the M{\"o}bius
strip ($\M$). It takes the generic form:
\be
Z_{1-\rm loop}={1\over 2}(\T+\K)+{1\over 2}(\A+\M)\ ,
\label{pf}
\ee
where the two terms in brackets correspond to closed (untwisted) and open
(twisted) states. In particular,
$\K$ and $\M$ describe the propagation of closed and open unoriented strings, 
imposing the $\Omega$ projection in the two sectors of the theory.

T-duality can be incorporated into the open string framework with the main
effect of interchanging Neumann (N) with Dirichlet (D) boundary conditions.
In fact, imposing NN conditions in some compact direction $X\equiv X+2\pi R$
yields
\be
X(\sigma,\tau)= X_L(\tau+\sigma)+X_R(\tau-\sigma)\, ,
\label{X}
\ee
and gives the expression (\ref{mode}) with the quantized momentum $p=n/R$.
In analogy with closed strings,
T-duality transforms $X_L\to X_L$, $X_R\to -X_R$, and inverts
the radius $R$ to $\tilde R=\alpha'/R$ (\ref{R1R}). 
It follows that $X\to{\tilde X}$, with the dual coordinate
\ba
\tilde{X}(\sigma,\tau)&=& 
X_L(\sigma+\tau)-X_R(\sigma-\tau)\, ,\label{Xtil}\\
&=&\tilde{x} +2nR \sigma 
  +i\sqrt{2\alpha'}\sum_{l\neq 0}{\alpha_l\over
  l}\sin l\sigma\, .\nonumber 
\ea
We see that $\tilde{X}(0,\tau)=\tilde{x}$ and
$\tilde{X}(\pi,\tau)=\tilde{x}+2\pi nR$, which means that 
the string is wrapped $n$ times around the circle and has
Dirichlet boundary conditions on a D-brane at $\tilde{X}=\tilde{x}$.
Thus a Dp-brane compactified on a circle is exchanged under T-duality with a
D(p-1)-brane obtained by wrapping around the compact dimension.

{}Furthermore, under T-duality, $\Omega$ is mapped to
$\tilde{\Omega}=\Omega{\cal R}$ where ${\cal R}$ is the ${\bf Z}_2$ orbifold
transformation $X \to -X$. Indeed, starting from Neumann boundary conditions for
the compact coordinate $X$ (\ref{X}) and performing a T-duality
transformation, one finds that the dual coordinate $\tilde X$ (\ref{Xtil})
is antiperiodic under the exchange of left- and right-movers.

{}Finally, T-duality exchanges Wilson lines of the gauge fields associated with
Neumann boundary conditions with the D-brane positions in the compactified space.
In fact, in the presence of Wilson lines $a_{i,j}$ corresponding to an
NN (traceless) gauge group generator $T_{(i,j)}={\rm diag}
(0\dots {1_i}\dots {{-1}_j}\dots 0)$, on a circle, the compact momentum is
shifted as:
\be
{n\over R}\to {n+a_i-a_j\over R}\ .
\label{wl}
\ee
Using the same
line of reasoning as before, it is easy to see that the dual coordinate 
satisfies ${\tilde X}(\pi,\tau)-{\tilde X}(0,\tau)=2\pi (n+a_i-a_j)R$.
This shows that up to a common shift, $a_i$'s represent positions of the open
string ends, or equivalently the location of D-branes.

\section{EFFECTIVE FIELD THEORIES}

Effective Lagrangians, and most of all
their dependence on the dilaton fields, provide a powerful guide
into various duality symmetries.
Let $\phi_n$ denote the dilaton field in $n$ non-compact dimensions.
Focusing on the Einstein and Yang--Mills 
kinetic terms of the
heterotic $SO(32)$ and type I string effective Lagrangians we have
\be
{\cal L}= e^{-2\phi_{10}}\left\{ {1\over 2}R+{1\over
      4} F^2\left| \matrix{\hspace{-.3cm} 1 \cr
        e^{\phi_{10}}}\right.+\dots
    \right\} ,\label{herve}
\ee
where the heterotic and type I gauge kinetic factors are assembled
together as upper and lower entries, respectively.
These Lagrangian terms contain also implicit dimensionful factors,
which are all equal to 1 in the units of $\alpha'$.

After compactification from 10 to $D\leq 10$ dimensions on
the same manifold of volume $V_{10-D}$,
\be
{\cal L}= e^{-2\phi_{10}}V_{10-D}\left\{ {1\over 2}R+{1\over
      4} F^2\left| \matrix{\hspace{-.3cm} 1 \cr
        e^{\phi_{10}}}\right.+\dots 
    \right\}\, .
\ee
The above expressions allow the following identification of dilatons 
in dimension $D$:
\be
e^{-2\phi_D}\equiv e^{-2\phi_{10}}V_{10-D}\, ,
\label{dil}
\ee
in terms of which we have
\be
{\cal L}= e^{-2\phi_{D}}\left\{ {1\over 2}R+{1\over
      4} F^2\left| \matrix{\hspace{-1cm} 1 \cr
        e^{\phi_{D}}V^{1/2}_{10-D}}\right.+\dots\right\} .
\ee
{}Finally, we go to the Einstein frame by rescaling the metric
$g_{\mu\nu}\to g_{\mu\nu}e^{4\phi_D/(D-2)}$ to obtain
\be
{\cal L}={1\over 2}R+{1\over
      4} F^2\left| \matrix{\hspace{-1cm} e^{{-4\over D-2}\phi_D} \cr
        e^{{D-6\over D-2}\phi_{D}}V^{1/2}_{10-D} }\right.+\dots 
\label{CN}
\ee

For $D=10$, the identification of
the two Lagrangians implies $\phi_{10}^H=-\phi_{10}^I$; the two string theories
are equivalent under S-duality.  In each case, since the
gravitational coupling constant is
$\kappa_{10}^2=e^{2\phi_{10}}\alpha'^4$ \cite{Wvar,PW},
\be
\lambda_{10}\leftrightarrow 1/\lambda_{10}\quad ; \quad
\alpha'\leftrightarrow\lambda_{10}\alpha'\, . 
\label{conj10} 
\ee

A derivation of type I and heterotic theories from eleven dimensions
provides more insight into their duality. One can see that M-theory
compactified on  $S^1/{\bf Z}_2\times S^1$ and $S^1 \times S^1/{\bf Z}_2$
gives rise to the heterotic and type I strings compactified on $S^1$ \cite{HW}.
Here, ${\bf Z}_2$ acts as the reflection on the eleventh and tenth 
coordinate, respectively,
changing also the sign of the 3-form. Moreover in the second case, one can show 
that the ${\bf Z}_2$ symmetry is identical to $\Omega \R$ which, as shown in
Section 2, is T-dual to the world-sheet involution of type IIB, allowing
construction of type I strings from type IIA. 
The corresponding identifications are \cite{HW}:
\ba
\phantom{.}&\hspace{-.8cm}&\lambda^I ={R^M_{11}\over R^M_{10}}=1/\lambda^H
\, ,\label{TI}\\
\phantom{.}&\hspace{-.8cm}&R^I_{10}={1\over R^M_{10}\sqrt{R^M_{11}}}\quad ,
\quad
R^H_{10} ={1\over R^M_{11}\sqrt{R^M_{10}}}\, ,\nonumber \\
\ea
where $R^M_{10,11}$, $R^I_{10}$ and $R^H_{10}$ are the compact radii in the
respective theories. 
Thus, S-duality is now recovered from U-dualities of
M-theory compactified on a torus
whose complex structure (\ref{TI}) is identified with the coupling constants. 

For $D=6$, one derives from Eq.\  (\ref{CN}) a U-duality relation
$e^{-2\phi_6^H}= V_4^I$. Moreover, using Eqs.\ (\ref{dil}) and
(\ref{conj10}), one obtains
the same identification with the indices $H$ and $I$ exchanged, hence
\be
V_4 \leftrightarrow 1/\lambda_{6}^2\, .
\label{conj6}
\ee 

For $D=4$, Eq.\ (\ref{CN}) implies a weak--weak coupling duality with
$e^{-2\phi_4^H}=e^{-\phi_4^I}{(V_6^{I})}^{1/2}$. Using Eqs.\ (\ref{dil}) and
(\ref{conj10}), one also obtains 
$V_6^H=e^{-3\phi_4^I}{(V_6^I)}^{-1/2}$ and similar relations with
the indices $H$ and $I$ exchanged. Finally, we get \cite{abt}
\be
\lambda_4^2\leftrightarrow \lambda_4 V_6^{-1/2}\quad ; \quad V_6
\leftrightarrow \lambda_4^{-3}V_6^{-1/2}\, .
\label{conj4}
\ee

It is interesting that the ten-dimensional strong--weak coupling duality between
heterotic and type I strings leads in lower dimensions to relations that hold
when both theories are weakly coupled and thus, can be tested at the
perturbative level. These relations can also be used to study
non-perturbative effects in one theory by using the perturbative description 
of the dual theory. For instance, world-sheet instanton contributions of the 
characteristic
magnitude $e^{-V_4^{1/2}/\alpha'}$ or $e^{-V_6^{1/3}/\alpha'}$ are mapped 
through the relations (\ref{conj6}) and (\ref{conj4}) to
non-perturbative effects of order $e^{-1/\lambda_{6}}$ and $e^{-1/\lambda_{4}}$,
respectively. Since $\lambda_4^H$ is also the gauge coupling, such world-sheet
instantons on the type I side translate on the heterotic theory to
stringy non-perturbative effects of order $e^{-1/g}$, which are much stronger
than ordinary field-theoretical instantons.

\section{TYPE I--HETEROTIC DUALITY IN SIX DIMENSIONS}

The first non-trivial test of heterotic--type I duality is for six-dimensional
vacua with $N=1$ supersymmetry, obtained upon compactification of the
ten-dimensional theories on $K3$ manifolds \cite{berk}. 

We start by recalling some basic
facts about $D=6$, $N=1$ supersymmetry. Besides the gravitational multiplet,
which contains 12 bosonic and 12 fermionic degrees of freedom, namely the
metric, an anti-self-dual 2-index antisymmetric tensor and the gravitino
(with 9, 3 and 12 physical components respectively),
there are three different matter multiplets:
the tensor ($T$), the vector ($V$) and the hyper ($H$),
containing $4+4$ physical states. These are a self-dual 2-form and a real
scalar (for $T$), a vector (for $V$) and two complex scalars (for $H$), together
with a Weyl spinor. Cancellation of the
gravitational anomaly in six dimensions
requires the number of massless matter multiplets $n_T$, $n_V$ and $n_H$ 
to satisfy the following constraint:
\be
29 n_T + n_H - n_V = 273\, .
\label{ac}
\ee

Perturbative heterotic spectra always contain one tensor multiplet whose scalar
component is identified with the six-dimensional dilaton $\phi_6^H$. All other
moduli belong to hypermultiplets, since vector multiplets contain no scalars.
Moreover, Eq.\ (\ref{ac}) fixes the difference between hypers and vectors,
$n_H - n_V = 244$. On the other hand, perturbative type I strings may contain
several tensor multiplets. Therefore in order to make perturbative duality tests
possible, we will restrict ourselves to type I vacua with $n_T=1$. 
In contrast to
the heterotic case, however, the type I dilaton $\phi_6^I$ belongs to a
hypermultiplet, in agreement with the duality relation (\ref{conj6}).

At a generic point of the moduli space, all massless hypermultiplets are neutral
under the gauge group. Supersymmetry requires that
hypermultiplets decouple from the interactions of vectors and
tensors, so that the effective action splits into two terms:
$S_{eff}=S_{T,V}+S_H$, where $S_{T,V}$ depends on tensors and vectors only,
while $S_H$ describes the interactions of hypers. The important point is
that the type I string coupling is given by a vev of a hypermultiplet, while the
heterotic coupling is determined by a vev of a tensor multiplet. It follows
that 
$S_{T,V}$ receives no quantum corrections on the type I side, while $S_H$ is
determined classically on the heterotic side.

Here, we focus only on $S_{T,V}$, which is easier to analyze. In fact, its form
is almost completely fixed by supersymmetry and anomaly cancellation. In the
Einstein frame, the bosonic part of the Lagrangian becomes \cite{s1}:
\begin{eqnarray}
{\cal L}_{T,V}^{(6)}&=&{1\over 2}R +{1\over 2}\left( {\partial\omega\over\omega}
\right)^2 +{1\over 4}(v\omega +v'\omega^{-1})F^2\nonumber\\
&+& \!\!\!{1\over 16}\omega^2(dB-v\Omega)^2 +{v'\over 16}B\wedge F\wedge F ,
\label{LTV}
\end{eqnarray}
where $\omega$ is the scalar of the tensor multiplet, $B$ is the 2-form
potential, $F$ denotes the field-strengths of gauge fields with $\Omega$ the
corresponding Chern--Simons terms, $v$ and $v'$ are gauge group-dependent
constants, and an implicit summation over gauge fields is understood.

On the heterotic side, $\omega\equiv e^{-\phi_6^H}$, therefore $v$ is given
classically by the Ka\v{c}--Moody level $k$ of the underlying affine Lie
algebra, while $v'$ represents a one-loop string correction. The
difference $\Delta v'$ between two gauge groups coincides with the corresponding
difference of $N=2$ $\beta$-functions that one obtains upon further
compactification to $D=4$ on a $T^2$ \cite{afiq}. This can be compared
with the result of the dual type I theory, where $\omega\equiv V^{1/2}_{K3}$, 
and both $v$ and $v'$ are determined at the tree level. 

Gauge fields with $v=0$ are non-perturbative from the heterotic point of view
and correspond to parallel 5-branes on the type I side. On the other hand, when
$vv'<0$ for some gauge field, there is a point in the tensor moduli space where
the gauge coupling blows up. It has been argued that such a singularity is 
resolved
by the appearance of tensionless strings \cite{SW3}, generated on the 
type I side by type IIB
3-branes wrapping around a collapsing 2-cycle of $K3$. Note finally
that if there is a $D=6$, $N=1$ type II vacuum with only one antisymmetric
tensor, $v'$ will be given at the tree level and $v$ at one loop, as required
by heterotic--type II S-duality in six dimensions.

We conclude this Section by describing one of the
simplest $N=1$ heterotic--type I dual pairs
in six dimensions. They are 
obtained by compactifying the ten-dimensional theories
on the $K3$ orbifold $T^4/\R$, where $\R$
is the ${\bf Z}_2$ transformation $X^m\to -X^m$ for $m=6,7,8,9$.

On the heterotic side, the ${\bf Z}_2$ projection reduces the supersymmetry to
$N=1$ and breaks the gauge group
$SO(32)$ down to $U(16)$. From the original $SO(32)$ gauge multiplet there
remain also two massless hypermultiplets transforming
in the ${\bf 120}_{1/2}$ representation. In addition, there are 4
untwisted neutral hypermultiplets parametrizing the moduli of $T^4$, as well as
16 twisted ones transforming in the ${\bf 16}_{1/4}$ representation. Of course,
there is also the standard $N=1$ supergravity multiplet together with the tensor
dilaton multiplet.

This model is dual to the type I theory compactified on the same $T^4/\R$ 
which can be constructed from type IIB by modding it out by the orientifold
group $\{ 1,\Omega, \R, \Omega \R\}$ \cite{bs,gp}. Then, besides the 16 D9--branes,
tadpole cancellation requires the presence of 16 D5--branes.
The massless spectrum of this $K3$ orientifold model is as follows. In the 
$\Omega$-twisted sector, open strings satisfy 99 boundary
conditions and the $SO(32)$ gauge group is broken by the $\R$ projection
to $U(16)_9$. In the $\Omega \R$-twisted sector, one finds
55 strings with both endpoints fixed on $T^4/\R$, and an additional 
5-brane gauge group of rank 16. The maximal gauge group is $U(16)_5$ which
appears when all 5-branes are located at the same fixed
point of the orbifold. There are also
two massless hypermultiplets transforming in the 
${\bf 120}_{1/2}$ representation of $U(16)_5$, 
which parametrize the relative distances
between the D5-branes, as mentioned in Section 2. When the 5-branes are pulled
apart the gauge group is broken, and the maximum breaking occurs when each
5-brane is located at a separate fixed point; the gauge group is then 
$U(1)_5^{16}$ and there is no massless matter. There are also
$95+59$ strings which give 16 massless hypermultiplets in the 
${\bf 16}_{1/4}$ representation of $U(16)_9$ [or 1 in the
$({\bf 16}_{1/4},{\bf 16}_{1/4})$ of $U(16)_9\times U(16)_5$]. Finally,
the closed string
sector contains the $N=1$ six-dimensional supergravity multiplet, one tensor
multiplet and 20 hypermultiplets, 4 of them from the $\R$-untwisted sector and
16 from the $\R$-twisted one. Out of the four scalar components of each 
twisted hypermultiplet,
three are coming from the symmetrized NSNS sector while the fourth one
originates by Poincar\'e duality from a RR 4-form potential that is equivalent 
to a scalar in six dimensions.

The above type I massless spectrum is the same as in the heterotic model with
the exception of the $U(1)_5^{16}$ gauge multiplets which are non-perturbative
on the heterotic side, and 16 hypermultiplets from the $\R$-twisted closed
string sector. However, all $U(1)_5$'s are broken due to anomalous couplings
of their gauge field strengths with the RR 4-forms, 
$C_4^i\wedge F^i_2$ \cite{d}. As a result
these gauge multiplets become massive, absorbing all 16 twisted hypermultiplets,
and the massless spectra of the two theories become identical. The effective
Lagrangian that describes the interactions of the 9-brane gauge multiplets
with the tensor multiplet is of the form (\ref{LTV}) with $v=1$ and $v'=0$.
In Section 7, we will compactify the above dual pair on
$T^2$ and discuss some non-trivial duality tests in four dimensions.

\section{TYPE I--HETEROTIC DUALITY IN FOUR DIMENSIONS}

When heterotic and type I theories are compactified on $K3\times T^2$, one
obtains $N=2$ supersymmetric models in four dimensions with $n_V+3$ vector
multiplets and $n_H$ hypermultiplets. The three additional vector multiplets
describe the universal sector which arises from the single tensor multiplet and
the Kaluza-Klein reduction of the graviton multiplet on $T^2$. 
More precisely, the vector field components
correspond to three linear combinations of $G_{\mu 4}$, $G_{\mu 5}$, $B_{\mu 4}$
and $B_{\mu 5}$, while the fourth combination can be identified with 
the graviphoton of
the $N=2$ supergravity multiplet. Similarly, their scalar 
components are combinations
of the three $T^2$ metric components $G_{44}$, $G_{55}$ and $G_{45}$, 
the internal
antisymmetric tensor $B_{45}$, the 
four-dimensional dilaton $\phi_4$, the $K3$ volume $V$, and the
universal axion $\alpha$ dual to the antisymmetric tensor $B_{\mu\nu}$.
In addition, there are $2n_V$ scalar components of the six-dimensional gauge
fields $a_{4,5}$ that correspond to the Wilson lines on $T^2$.
The way these fields are arranged into $N=2$ multiplets is different in
the two theories. 

We first discuss the case of vanishing Wilson lines.
On the heterotic side, the $K3$ volume belongs to a hypermultiplet while the
remaining six scalars form the well-known complex fields:
\ba
S_H=\alpha+ie^{-2\phi_4}&,& T=B_{45}+i\sqrt{G}\nonumber\\
&&\hspace{-2cm}U=( G_{45}+iG^{1/2}) /G_{44}\, ,
\label{stu}
\ea
where $G=G_{44}G_{55}-G_{45}^2$. 
On the type I side, the six-dimensional
string dilaton $\phi_6=\phi_4+(1/2)\ln\sqrt{G}$ remains in a hypermultiplet.
A straightforward dimensional reduction
of the six-dimensional Lagrangian (\ref{LTV}) gives \cite{abt}
\begin{equation}
{\cal L}_V^{(4)}{=} {1\over 2} R+{1\over 4}
(v\makebox{Im}S_I+v'\makebox{Im}S')F^2+\dots  \label{four}
\end{equation}
where 
\ba
S_I &=& \alpha+ie^{-\phi_4}G^{1/4}V^{1/2}\, ,\nonumber\\ 
S' &=& B_{45}+ie^{-\phi_4}G^{1/4}V^{-1/2}\ . 
\label{s}
\ea
It is easy to see that with the above definitions and with $U$ 
the same as in Eq.\  (\ref{stu}), 
$S_I$, $S'$ and $U$ 
have diagonal kinetic terms and form the scalar components of
the three vector multiplets.

As seen in Eq.\ (\ref{stu}) the heterotic dilaton belongs to a vector multiplet,
which implies that the hypermultiplet moduli space remains the same as in six
dimensions and does not receive any quantum corrections.
On the other hand, on the type I side, the four-dimensional
dilaton $\phi_4$ is a ``mixture'' of vector and hypermultiplet
components, so that the string coupling $e^{\phi_4}=
e^{\phi_6}(\mbox{Im}S_I\mbox{Im}S')^{-1/4}$.
Hence, both hyper and vector sectors can
receive quantum corrections once the string coupling combines
with other fields to form the full scalar supermultiplet components.

Wilson lines can be turned on along the flat directions of the potential
corresponding to the Cartan directions of the gauge group.
They can be diagonalized to $a_{4,5}^i$, $i=1\dots r$, 
where $r$ is the rank,
and generically break the gauge group to its abelian Cartan subgroup $U(1)^{r}$.
In the presence of Wilson lines, 
the $N=2$ special coordinates for the type I theory become:
\begin{eqnarray}
A_i=a^i_4-a^i_5U\, ,&& \hspace{-.5cm}S_I=S_I|_{A=0}+\sum_i{v'_i\over 2} a^i_5
A_i\, ,\nonumber\\
 && \hspace{-.5cm}S'=S'|_{A=0}+\sum_i{v_i\over 2} a^i_5 A_i\, .
\label{S'redef}
\end{eqnarray}
Similarly, in the heterotic theory, $A_i$ is defined as above, 
the $T$ modulus is redefined as $S'$, while $S_H$ remains the
same as in Eq.\  (\ref{stu}).

In terms of these fields, the tree-level
prepotentials that determine the interactions of $N=2$ vector multiplets 
in the two theories can be read off from the six-dimensional action (\ref{LTV})
and take the form \cite{abt}:
\ba
{\cal F}^I_{\mbox{\scriptsize
    tree}} &=& S_IS'U-\frac{1}{2}\sum_i(v_iS_I+v'_iS')A_i^2\nonumber\\
{\cal F}^H_{\mbox{\scriptsize tree}} 
&=& S_HTU-\frac{1}{2}\sum_i v_iS_H A_i^2\, .
\label{preps}
\ea
It follows that the heterotic--type I duality mapping
inherited from Eq.\  (\ref{conj6}) in $D=6$, is \cite{abt}
\begin{equation}
S_H\leftrightarrow S_I,~~ T\leftrightarrow S',~~ U\leftrightarrow U,~~
A_i\leftrightarrow A_i\, .
\label{mapf}
\end{equation}

Let us now discuss quantum corrections. As already mentioned above, both
prepotentials can be modified quantum mechanically. Perturbative corrections
are restricted by continuous Peccei--Quinn symmetries. On the heterotic side,
there is a Peccei--Quinn symmetry associated with the universal axion
$\alpha={\rm Re}S_H$, which is dual to the antisymmetric tensor.
On the type I side, besides a similar symmetry associated with
Re$S_I$, there is a second symmetry due to the fact that Re$S'=B_{45}$
originates from the RR sector. These symmetries, together with
analyticity, imply that the perturbative corrections must be independent
of $S_{I,H}$ and $S'$, hence they are entirely due to one-loop effects.

The general forms of the respective prepotentials are:
\begin{eqnarray}
{\cal F}^I(S,S',U,A)\!\!\! &=& \!\!\!{\cal F}^I_{\mbox{\scriptsize tree}}+
f^I(U,A)+\dots\nonumber\\
{\cal F}^H(S,T,U,A)\!\!\! &=& \!\!\!{\cal F}^H_{\mbox{\scriptsize
tree}}+f^H(T,U,A)+
\dots\label{fper}
\end{eqnarray}
where $f^{I,H}$ are the one-loop corrections and dots
represent non-perturbative contributions. 
In Eq.\  (\ref{fper}) and everywhere below we drop the subscripts I and H 
referring to $S$, in view of the identification (\ref{mapf}).
Since the Peccei--Quinn symmetries are broken by instanton effects
to discrete shifts of axions, the non-perturbative corrections
can be expanded in integer powers of $e^{2\pi iS}$ whose magnitude is suppressed
by the instanton action.

As already mentioned before, heterotic--type I duality includes
regimes that are weakly coupled on both sides.
It is possible to compare prepotentials 
(as well as other quantities that will be discussed later) already
at the perturbative level, in the appropriate limits of moduli
parameters. First of all, in order to reach the perturbative
limit on type I side, one has to take the limit of large
Im$S$ and Im$S'$. This is mapped via  Eq.\  (\ref{mapf}) to the region
of large Im$S$ and Im$T$ which corresponds to a weakly coupled
heterotic theory in the limit of ``large'' $T^2$ torus compactification,
provided that the limit is taken with Im$S>{\rm Im}S'$.
Thus the perturbative heterotic prepotential reproduces its type I counterpart
in the region of $K3$ volume 
$V>1$,  cf.\ Eq.\  (\ref{s}). More precisely, from the expansions (\ref{fper}),
one obtains
\begin{equation}
f^H(T,U,A)\stackrel{T\to i\infty}{\longrightarrow}\! -{1\over 2}\sum_i
v'_i\, TA_i^2 +f^I(U,A), 
\label{pretest}
\end{equation}
with the fields mapped as in Eq.\  (\ref{mapf}). Note that a part of the 
one-loop heterotic prepotential is mapped
into the tree-level type I term proportional to $v'_i$,
while the remaining part reproduces the type I one-loop correction.

It is also interesting to study the other perturbative branch of the type I 
theory with $V<1$, which 
is mapped in the heterotic theory to the non-perturbative region 
Im$T>{\rm Im}S\to\infty$. These two regions are related by type I T-duality
$V\rightarrow 1/V$ which corresponds on the heterotic side to the
non-perturbative $S\leftrightarrow T$ exchange.
If in addition there exists a type IIA description of the same model, the exact
prepotential can be determined classically on the type II side, and 
this region can be probed directly, providing a perturbative
test of type I--type II duality
that is non-perturbative from the heterotic point of view.

\section{ONE-LOOP CORRECTIONS}

The standard way of deriving the one-loop
prepotential in heterotic theory \cite{agnt,prep} 
is to extract it from the one-loop
corrections to gauge couplings:
\begin{equation}
\frac{4\pi^2}{g^2}={\pi\over 2}\makebox{Im}S+\Delta^H\, ,
\end{equation}
where $\Delta^H$ is the threshold function \cite{dkl}.
Its moduli dependence is governed by  \cite{agnt}
\begin{equation}
\partial_U\partial_{\bar{U}}\Delta^H=bK^{H(0)}_{U\bar{U}}+4\pi^2
K^{H(1)}_{U\bar{U}},  \label{delhet}
\end{equation}
where $b$ is the beta-function coefficient, 
$K^{H(0)}_{U\bar{U}}=-1/(U-\bar{U})^2$ is the tree-level moduli metric
and $K^{H(1)}_{U\bar{U}}$
is the one-loop correction.
The first term in the r.h.s. of Eq.\  (\ref{delhet}) depends on the gauge group
while the second one
is a universal, gauge-group--inde\-pendent contribution.
The latter can be used
to extract the one-loop prepotential $f^H$.

A direct string computation gives $\Delta^H$ as an integral over the complex 
Teichm\"uller parameter $\tau=\tau_1+\tau_2$ of the world-sheet torus in its
fundamental domain \cite{agnt}:
\be
\Delta^H\! =-{1\over 2} \int {d^2\tau\over\tau_2} 
\bar{\eta}^{-2} {\rm Tr}_R F(-)^F
\big(Q^2-{1\over 2\pi\tau_2}\big),
\label{del}
\ee
where $\eta$ is the Dedekind eta function, $Q$ is the gauge group generator, and
the trace is over the Ramond sector of the internal $(2,0)$ superconformal
theory with $U(1)$-charge operator $F$. After taking 
$\partial_U\partial_{\bar U}$ 
derivatives, one finds that the two terms in the trace give rise to the two
terms in the r.h.s.\ of Eq.\  (\ref{delhet}), respectively. Using the 
properties of
the underlying superconformal field theory that describes $K3\times T^2$
compactifications, one can rewrite the integrand as a sum over $N=2$ BPS
states \cite{hm}. In particular, the one-loop K\"ahler metric reads:
\ba
K^{H(1)}_{U\bar{U}}&=& -{1\over 8\pi^2}\int {d^2\tau\over\tau_2}
\partial_U\partial_{\bar U}\label{bpshet}\\
&&\hspace*{-1.8cm}
\bigg( \sum_{\rm BPS\atop\rm hypermultiplets}-
\sum_{\rm BPS\atop\rm vectormultiplets}\bigg)
e^{i\pi\tau M_L^2}e^{-i\pi\taubar M_R^2}\ ,\nonumber
\ea
where $M_L$ and $M_R$ denote the contribution to the masses from the left- and
right-movers, respectively. In this formula, the supergravity multiplet is 
counted as a vector multiplet.

One-loop threshold corrections have also been studied
in type I theory \cite{bf}, however their structure is different
from the heterotic case. The function $\Delta^I$ contains
the group-dependent contribution only, proportional
to the beta function.
The universal term is absent, which means that it is automatically
absorbed into the definition of Im$S$.
Hence another procedure is needed to compute the one-loop
K{\"a}hler metrics. It turns out that the quantity
to be examined is the Planck mass \cite{abt}.
Unlike the heterotic case, the Einstein term receives
a one-loop correction:
\begin{equation}
e^{-2\phi}R\stackrel{\makebox{\footnotesize 1-loop}}
{\longrightarrow} \big( e^{-2\phi}+{\delta\over\sqrt{G}}\big) R.
\end{equation}
The relation between the function $\delta$ and the  K{\"a}hler metric
is\footnote{The present definition of $\delta$ differs from its original
definition in Ref.\ \cite{abt} by a factor of $\sqrt{G}$, so that $\delta$ 
becomes independent of $\sqrt{G}$.}:
\begin{equation}
K^{I(1)}_{U\bar{U}}=
{1\over 16\pi \mbox{Im}S'}\partial_U\partial_{\bar{U}}\delta\,.
\label{kaler}
\end{equation}

The computation of the one-loop correction to the Planck mass
in type I theory has been presented in \cite{abt}. The function
$\delta$ can be determined from a physical amplitude
involving one modulus and two gravitons, and receives contributions
from the annulus and M{\"o}bius strip
diagrams for open strings, and from the Klein bottle for
closed strings. The result can be expressed as an integral over the real modular
parameter of these one-loop surfaces:
\be
{\delta\over\sqrt{G}} = -{1\over \pi}
\int_0^{'\infty}\!\frac{dt}{t^2}
(\I_{\A}Z_{\A}+\I_{\M}Z_{\M}+4\I_{\K}Z_{\K}),
\label{Is}
\ee
where $\I_{\sigma}$ are indices associated to $K3$ which count the open
string spinors (propagating on $\sigma=\A,\,\M$) and closed
string RR bosons (propagating on $\sigma=\K$) weighted with the 
fermion-parity operator
$(-)^{F_{\mbox{\tiny int}}}$, while $Z_{\sigma}$ denote the corresponding $T^2$
partition functions. The prime in the integral indicates that the quadratic
divergence in the ultraviolet limit $t\to 0$ has been subtracted, as
dictated by the tadpole cancellation \cite{bf}. There is no need however for
such a regularization at the level of the K\"ahler metric, where the
divergence disappears after taking $\partial_U\partial_{\bar{U}}$
derivatives, c.f.\ Eq.\  (\ref{kaler}).

The above result for the K\"ahler metric generalizes
the heterotic expression obtained
from Eq.\  (\ref{del}) to the type I case.
It can be reexpressed in a form similar to (\ref{bpshet}) 
as a sum over $N=2$ BPS states that originate only 
from the massless modes in six dimensions \cite{abt}:
\begin{eqnarray}
{1\over\sqrt{G}}\partial_U\partial_{\bar U}\delta  &=& 
-{2\over\pi}\int_0^{\infty}\frac{dt}{t^2}
\partial_U\partial_{\bar U}\label{bps}\\
&&\hspace*{-1.8cm}
\bigg( \sum_{\rm BPS\atop\rm hypermultiplets}-
\sum_{\rm BPS\atop\rm vectormultiplets}\bigg)
e^{-\pi t M^2/2}\, ,\nonumber
\end{eqnarray}
where the masses $M$ come from the momentum in the internal $T^2$.
The same formula gives type I threshold corrections after inserting the operator 
$tQ^2$ inside the sum.

\section{EXAMPLE OF TYPE I--HETERO\-TIC DUALITY}

In this Section, we discuss an example of a dual pair with $N=2$ supersymmetry
in four dimensions, obtained by compactifying on $T^2$ the six-dimensional
$K3$ orbifold models described in Section 4. More precisely, we will analyze
two cases. The first one corresponds to the Higgs phase, in which the 
six-dimensional gauge group $U(16)_9$ is completely broken by giving appropriate 
vev's to the charged hypermultiplets. One thus obtains $n_T=1$, $n_V=0$ 
and $n_H=244$ consistently with the anomaly cancellation condition
(\ref{ac}) in $D=6$. Upon compactification to $D=4$ one finds the so-called
STU model which contains 3 massless vector multiplets ($S$, $T$ and $U$ in the
heterotic notation) and 244 neutral 
hypermultiplets. The second case corresponds to the Coulomb phase 
obtained by turning on Wilson lines on $T^2$. $U(16)_9$ is then broken to
$U(1)^{16}$ and all charged hypermultiplets become massive. The resulting 
four-dimensional massless spectrum contains 19 vector multiplets and 4 neutral
hypermultiplets.

\subsection{Higgs Phase}

On the heterotic side, the STU model has been studied extensively in the past
because it admits also a type II description \cite{kv}. 
In fact, it can be obtained by
compactifying the ten-dimensional type IIA theory on the Calabi--Yau manifold
described by the weighted hypersurface of degree 24, $WP_{1,1,2,8,12}(24)$,
with Hodge numbers $h_{(1,1)}=3$ and $h_{(1,2)}=243$. Since the type II dilaton
belongs to a hypermultiplet, on the type II side the
exact prepotential can be determined at the classical level. Therefore, this
model provides an example of type II--heterotic--type I triality which can
be tested in appropriate limits at both 
perturbative and non-perturbative levels.

Starting from the type II prepotential $\F^{II}$ with the field identification
of $S,T,U$ guided by the $K3$ fibration, and taking the limit Im$S\to\infty$
one can perform a perturbative test of heterotic--type II duality
\cite{kv,checks}:
\be
\F^{II}(S,T,U)~\stackrel{S\to i\infty}{\longrightarrow} STU+f^H(T,U)\, ,
\label{II}
\ee
where the two terms in the r.h.s.\ coincide with the tree-level and the one-loop
contributions to the heterotic prepotential. A non-perturbative test 
\cite{kv} can also be
done by taking the zero-slope limit along the conifold singularity 
of the Calabi--Yau manifold, reproducing
the Seiberg--Witten prepotential of the rigid $SU(2)$ $N=2$ super-Yang--Mills
theory whose perturbative limit is described by the heterotic model on the $T=U$
line of enhanced symmetry.

In order to  test heterotic--type I duality, we first recall that
the heterotic $S$, $T$ and $U$ fields are mapped to $S$, $S'$ and
$U$ defined in Eq.\  (\ref{s}). As explained in Section 5, a perturbative
test can be performed by taking the limit of Im$T\to\infty$ of the one-loop
heterotic prepotential and comparing it with the type I counterpart.

The one-loop type I prepotential can be reconstructed from the K\"ahler
metric given by Eqs.\ (\ref{kaler},\ref{bps}).
In the model under consideration, Eq.\  (\ref{bps}) yields
\ba
\partial_U\partial_{\ubar}\delta&=&-{2\sqrt{G}\over \pi}\int_0^{\infty} {dt\over
  t^2}\,(244-4)\times\nonumber\\& &~~~~~\partial_U
  \partial_{\ubar}\sum_{p\in\Gamma_2}
  e^{-\pi t|p|^2/2}\, ,\label{bpstu}
\ea
where $\Gamma_2$ is the $T^2$ momentum lattice: 
\be
p={m_4-m_5 U\over \sqrt{2G^{1/2}\mbox{Im}U}}\, ,
\label{momI}
\ee
with integer $m_{4,5}$. After performing the $t$-integration, 
Eq.\  (\ref{kaler}) gives \cite{abt}:
\be
K^{I(1)}_{U\ubar}=-\frac{15}{\pi^4}\frac{1}{\makebox{Im}S'}{\sum_{m_4,m_5}}'
\frac{1}{|m_4-m_5 U|^4}\label{gloop}\ ,
\ee
where the prime means $(m_4,m_5)\neq (0,0)$.

The one-loop prepotential can now be determined by the standard $N=2$ 
formula \cite{prep}:
\ba
\partial_U^3f^I(U)&=&2\,{\rm Im}S'\,\partial^2_U\partial_{\ubar}(U-\ubar)^3
K^{I(1)}_{U\ubar}\nonumber\\&=&4\,E_4(U)\ ,\label{fiu}
\ea
with the function
\be
E_4(U)=\frac{45}{\pi^4}{\sum_{m_4,m_5}}'\frac{1}{(m_4-m_5U)^4}\ .
\ee
This result agrees with the heterotic prepotential \cite{prep} 
in the limit Im$T\to\infty$:
\be
\partial_U^3f^H(U,\makebox{Im}T\rightarrow\infty)=4\,
E_4(U)=\partial_U^3f^I(U)\ .
\label{eq}
\ee

Note that the perturbative type I computation is valid not only in the region
Im$S>{\rm Im}S'\to\infty$ ($K3$ volume $V>1$), but also in the region
Im$S'>{\rm Im}S\to\infty$ ($V<1$). This is due to
the symmetry of the present type I model under
T-duality, $V\leftrightarrow 1/V$, which
exchanges 9-branes and 5-branes. Although the region of $V<1$ cannot be reached
by means of perturbative expansion from the heterotic side, it can
be reached from the type II side by taking the limit Im$T\to\infty$ before
the limit
Im$S\to\infty$. Since the exact prepotential of the STU model, as evaluated
on the type IIA side, is
known to be symmetric under the exchange of $S$ and $T$ \cite{kv}, 
such an order of
limits gives $\F^{II}\to STU+f^I(U)$, the same as in the other order.

\subsection{Coulomb Phase}

In the Coulomb phase, we find it more convenient to test duality by 
a direct examination
of threshold corrections to the gauge couplings of $U(1)^{15}$ associated with
the unbroken Cartan generators of $SU(16)_9$ \cite{apt2}. After combining 
Eqs.\ (\ref{delhet}) and (\ref{kaler}) with
Eq.\  (\ref{pretest}), we see that duality predicts the following
large Im$T$ expansion of the heterotic threshold corrections:
\ba
\Delta^H(T,U,A)&=&{\pi\over 2}v'\mbox{Im}T+
\Delta^I(U,A)\nonumber\\
&&+\,{\pi\over 4\mbox{Im}T}\delta(U,A)+\dots\, ,
\label{ch}
\ea
up to exponentially suppressed corrections. Recall from the discussion of the
six-dimensional model in Section 5 that $v'=0$.

The type I quantities $\Delta^I(U,A)$ and $\delta(U,A)$ have been discussed 
for generic models in Section 6. In order to apply Eqs.\ (\ref{Is}) 
and (\ref{bps}) 
to the model under consideration, we first determine the quantities
$\I_{\sigma}Z_{\sigma}$ for various surfaces $\sigma=\A,\M,\K$. 
In the 99 open string sector,
\ba
\I_{\A}Z_{\A}&=&-2\!\!\!\!\!\sum_{a^I+a^J+\Gamma_2}
\!\!\!\!\!s_{IJ}e^{-\pi t|p|^2/2}\nonumber\\
\I_{\M}Z_{\M}&=&-2\!\!\!\sum_{2a^I+\Gamma_2}
\!\!\!\!\!e^{-\pi t|p|^2/2}\ ,
\label{ia1}
\ea
where for convenience, we introduced the
index $I\equiv i$ or $\ibar$, with $i$ and $\ibar$ running over the ${\bf
16}$ and ${\bf \overline{16}}$ of $SU(16)$, respectively, and $a^{\ibar}\equiv
-a^i$. The matrix $s_{IJ}$ represents the action of the orbifold group $\R$ 
on the Chan--Paton charges: $s_{IJ}=-1$ or 1, depending on whether $I$ and $J$ 
belong to the same or conjugate representations of $SU(16)$, respectively.
For open string surfaces with boundaries, the momenta of the $T^2$ lattice 
(\ref{momI}) are shifted in a self-explanatory way by the Wilson lines of the
9-brane group, according to Eq.(\ref{wl}); for instance:
\be
a^I+\Gamma_2\, : m_{4,5}\to m_{4,5}+a^I_{4,5}\, .
\ee
Similarly, in the 95+59 sectors:
\be
\I_{\A}Z_{\A}=2\times 16\!\!\!\sum_{a^I+\Gamma_2}
\!\!e^{-\pi t|p|^2/2}\, ,~~\I_{\M}=0\, .
\label{ia2}
\ee

There should be no contribution from the 55 and closed string sectors since 
the 5-brane $U(1)^{16}$ gauge supermultiplets 
become massive by absorbing  16 twisted 
hypermultiplets from the closed string sector, as explained in Section 4.
However, for a generic position of  5-branes, the 55 sector
contributes
\be
\I_{\A}Z_{\A}+\I_{\M}Z_{\M}=-4\times 16\sum_{\Gamma_2}
e^{-\pi t|p|^2/2}\, ,
\ee
while the closed string sector gives
\be
\I_{\K}Z_{\K}= 16\sum_{\Gamma_2}
e^{-\pi t|p|^2/2}\, ,\label{lastt}
\ee
due to the RR components of the twisted hypermultiplets. Note that these two 
contributions cancel each other.

Inserting the above results (\ref{ia1}-\ref{lastt}) into
Eq.\  (\ref{Is}), one obtains \cite{apt2}:
\ba
\delta &=& {2\sqrt{G}\over \pi}
\int_0^{'\infty}\frac{dt}{t^2}
\label{deltares}\\ 
&&\hspace{-0.9cm}\left\{ \sum_{a^I+a^J+\Gamma_2}
\!\!\!\!\!s_{IJ}\;
+\sum_{2a^I+\Gamma_2} -\; 16\sum_{a^I+\Gamma_2}
\right\} e^{-\pi t|p|^2/2}\, .\nonumber
\end{eqnarray}
A similar formula, obtained by inserting the charge operator $tQ^2$ in the sum, 
gives the type I threshold corrections \cite{bf}:
\ba
\Delta^I(U,A_i)= -{1\over 8}\int_0^{'\infty}\!{dt\over t} \left\{ 
\sum_{a^I+a^J+\Gamma_2}
\!\!\!\!\!s_{IJ}(q_I+q_J)^2+ \right. &&\nonumber
\end{eqnarray}
\be
\hspace{.5cm}\left. \sum_{2a^I+\Gamma_2}(2q_I)^2 -\; 16\sum_{a^I+\Gamma_2}q_I^2
\right\} e^{-\pi t|p|^2/2},
\label{DHl}
\ee
where $q_i$ are the $U(1)$ charges associated to SU(16) Cartan generators
in the $\bf 16$ representation and $q_{\ibar}=-q_i$.

We now turn to the heterotic side and determine the large Im$T$ behavior of 
$\Delta^H$ given by the general formula of Eq.\  (\ref{del}). Specifying 
to the model under consideration:
\begin{eqnarray*}
\Delta^H(T,U,A)= -{1\over 8} \int {d^2\tau\over \tau_2}
\sum_s Z_s(\taubar)\, \times
\end{eqnarray*}
\be
\sum_{p_L,p_R\in\Gamma_s^{(2,18)}}\left(Q^2-{1\over 2\pi\tau_2}\right)
e^{i\pi\tau|p_L|^2}e^{-i\pi\taubar|p_R|^2}\! ,\label{dell}
\ee
where the first sum is over the $N=2$ sectors of the $T^4/{\bf Z}_2$ orbifold,
$s=(P,A),(A,P),(A,A)$, where $P$ and $A$ denote periodic and antiperiodic 
boundary conditions, respectively.
$Z_s$ are the corresponding partition functions for all right-movers except for
the 18 right-moving momenta that are included in the $\Gamma_s^{(2,18)}$ 
lattices. In fact, the moduli dependence is due entirely to these lattices. In
particular, the left-moving (complex) momenta are:
\be
p_L={1\over{\sqrt{2\mbox{Im} T\mbox{Im}U}}}
(m'_4-m'_5U-n_5T-n_4TU) ,
\label{mom22}
\ee
with integer $n_{4,5}$ and the momentum numbers shifted appropriately
to $m'_{4,5}$ by the 16 Wilson lines. 

The limit Im$T\to\infty$ can be taken in the following steps \cite{apt2}. 
First, one observes that the contributions of all winding modes are 
exponentially
suppressed, hence one can restrict the lattice summation to $n_4=n_5=0$.
Next, by making the change of variable 
\be
\tau_2={t\over 4}\mbox{Im}T\ ,
\label{t}
\ee
one can
easily show that the integration domain becomes the strip 
$t\ge 4/{\rm Im}T$, $-1/2\le\tau_1\le 1/2$, up to exponentially small 
corrections in 
$\mbox{Im}T$. Finally, the $\tau_1$ integration selects the states that
originate from the massless modes only in $D=6$.

The leading contribution diverges linearly in Im$T$, reflecting the quadratic
ultraviolet divergence of the integral in the region $t\to 0$, where Im$T$ acts
as a regulator. The coefficient of the divergence determines $v'$ of
Eq.\  (\ref{ch}). It 
is given by the leading term of the expression (\ref{dell})  Poisson resummed
in the $T^2$ momentum numbers, with the $\Gamma_s^{(2,18)}$ lattices
degenerating to the moduli-independent $\Gamma_s^{16}$:
\ba
v'&=& -{1\over 4\pi}\int {d^2\tau\over \tau_2^2}
\sum_s Z_s(\taubar)\nonumber\\
&&~~~\sum_{p_R\in\Gamma_s^{16}}
\left(Q^2-{1\over 2\pi\tau_2}\right)e^{-i\pi\taubar|p_R|^2}
\label{vprime}
\ea
The above result coincides with the one-loop threshold correction 
of the six-dimensional heterotic theory. The integral (\ref{vprime}) 
can be shown
to vanish in the model under consideration, in agreement
with the fact that $v'=0$ on the type I side, as explained in
Section 4.

The subleading contribution is $T$-independent and comes entirely
from the $Q^2$ part of Eq.\  (\ref{dell}). 
It coincides with the type I expression for $\Delta^I$, Eq.\  (\ref{DHl}), after
using the condition $\sum_ia^i_{4,5}=0$ for the  Wilson lines corresponding
to $SU(16)$ Cartan generators.
Similarly, the term proportional to $1/\tau_2$ inside the bracket in
Eq.\  (\ref{dell}) reproduces the third term in Eq.\  (\ref{ch}), of order
$1/\mbox{Im}T$, with $\delta$ given in Eq.\  (\ref{deltares}).

It is interesting to trace  the origin of individual type I one-loop
contributions to the heterotic side. 
The contributions
of the 99 sector (on the annulus and M\"obius strip),
corresponding to the first two terms in Eqs.\ (\ref{deltares},\ref{DHl}),
originate on the heterotic side from the untwisted (P,A) orbifold
sector, while the contribution of the 95+59 sectors (on the annulus),
corresponding to the third term,
originate from the twisted (A,P)+(A,A) orbifold sectors.

\section{HIGHER-DERIVATIVE F-TERMS}

We now consider a class of higher-derivative F-terms in the effective
actions which, in $N=2$
superfield formalism, take the form 
\be
I_g=\F_gW^{2g}\ ,
\ee
with integer $g\ge 0$. Here, $W$ is the Weyl superfield
\be
W_{\mu\nu}=F^-_{\mu\nu} - R^-_{\mu\nu\lambda\rho}\theta^1
\sigma_{\lambda\rho}\theta^2+ \dots \ ,
\ee
which is anti-self-dual in the Lorentz indices. 
$F^-_{\mu\nu}$ and $R^-_{\mu\nu\lambda\rho}$ are the (anti-self-dual)
graviphoton field strength and Riemann tensor, respectively. 
The couplings $\F_g$ are
holomorphic sections of degree $2-2g$ of the vector moduli space, up to a
holomorphic anomaly for $g \geq 1$ \cite{BCOV}. 
They generalize the well known prepotential $\F\equiv \F_0$:
$\F_g$ $(g\geq 1)$ determines $(2g{+}2)$-derivative couplings 
of two gravitons and $(2g-2)$ graviphotons together
with all interactions related by supersymmetry \cite{top}.

On the type II side, $\F_g$ is determined entirely
at genus $g$, while on the heterotic and type I sides these couplings
arise at one loop (with additional tree-level contributions to
$\F_0$ and $\F_1$). The corresponding perturbative expansions in the two
theories are \cite{apt2}:
\begin{eqnarray*}
\F_1^H=4\pi\mbox{Im}S+f_1^H(T,U,A)\, ,
\end{eqnarray*}
\be
\F_g^H=\F_g^H(T,U,A)\qquad (g\geq 2)\, ,\label{FgH}
\ee
\begin{eqnarray*}
\F_1^I=4\pi(\mbox{Im}S+v'_1\mbox{Im}S')+f_1^I(U,A)\, ,
\end{eqnarray*}
\be
\F_g^I=\F_g^H(U,A)\qquad (g\geq 2)\, .\label{FgI}
\ee
In the example of dual pair discussed in Section 7, $v'_1=1$, as required by the
symmetry under type I T-duality.

The standard way of encoding $\F_g$'s is to combine them in the generating
function \cite{BCOV}:
\be
\F(\lambda)=\sum_{g=1}^\infty g^2\lambda^{2g}\F_g \, .
\label{gen}
\ee
Heterotic--type I duality predicts the following asymptotic expansion in the
large Im$T$ limit:
\ba
\F^H(\lambda;S,T,U,A)&=& 4\pi\lambda^2(\mbox{Im}S+\mbox{Im}T)\nonumber\\
&&\hspace{-3cm}
+\,\F^I(\lambda;U,A)
+\,{2\pi\lambda^2\over\mbox{Im}T}\delta(U,A)+\dots
\label{Fgexp}
\ea

The heterotic function $\F^H(\lambda)$ has been computed in Refs.\ 
\cite{AGNT,MS,apt2} and is given by an integral similar to (\ref{dell}) with the
operator inside brackets replaced by 
$-(\lambda^2/4\pi^2){d^2\over d\tilde{\lambda}^2} G^H(\tilde{\lambda},\tau)$,
where $G^H$ is the partition function of space-time coordinates in the presence
of a (anti-self-dual) graviphoton field strength background $\lambda$, and 
$\tilde{\lambda}\equiv{\bar{p}_L 
\tau_2 \lambda / \sqrt{2\, \mbox{Im}T \mbox{Im}U}}$.

The type I generating function $\F^I(\lambda)$ receives two types of
contributions. The first one can be obtained from the expression for the 
threshold corrections by a procedure similar to the heterotic case, and
arises from the world-sheet surfaces $\A$, $\M$ and $\K$. The second one arises
from the $N=4$ supersymmetric type IIB sector propagating on the world-sheet
torus; it contributes to $\F^I_1$ only. The sum of the two is \cite{MS2,apt2}:
\begin{eqnarray}
\F^I(\lambda)\!\!\!&\!\!\!\!=\!\!\!\!&\!\!\!{\lambda^2\over 64\pi^2}\!
\int_0^{'\infty}\!\! {dt\over t}
({\cal I}_{\A}Z_{\A}+{\cal I}_{\M}Z_{\M}+4{\cal I}_{\K}Z_{\K})\nonumber\\
&&\hspace{1.8cm}\times {d^2\over d\lambda^2}
\left( {\lambda \pi\over\sin\pi\tilde{\lambda}} \right)^2
\label{FIs}\\
&&+\, {\lambda^2\over 8}\int_0^{'\infty} {dt\over t}
({\cal I}_{\T}Z_{\T}-{\cal I}_{\K}Z_{\K})\ ,\nonumber
\end{eqnarray}
where the indices ${\cal I}_{\sigma}$ and partition functions $Z_\sigma$ have
been defined in Section 6, with the exception of ${\cal I}_{\T}$ and $Z_{\T}$
which are the Witten index (equal to 24 for $K3$) and the torus partition 
function, respectively. In contrast to the first integrand which
is reduced to a sum over $N=2$ BPS states as in Section 6, the 
second term $({\cal I}_{\T}Z_{\T}-{\cal I}_{\K}Z_{\K})$ receives
contributions only from the $\R$-untwisted $N=4$ sector of the theory.

Note that the integration in Eq.\ (\ref{FIs}) is infrared singular
at $t\to\infty$. The divergence is proportional to $\lambda^2$ and affects
$\F_1$ only, reproducing the trace anomaly of the effective field theory. 
Unlike the
case of gauge couplings, it is not regulated by non-vanishing Wilson lines. 

Specializing to the $K3$ orbifold model discussed in Section 7, 
in the Higgs phase we have:
\ba
\F^I(\lambda)\!\!&=&\!\!{\lambda^2\over 32\pi^2}\int_0^{'\infty}\!
 {dt\over t}\bigg\{240{d^2\over d\lambda^2}
\left( {\lambda \pi\over\sin\pi\tilde{\lambda}}\right)^2\nonumber\\
&&\;+\,{4\pi^2}(24-16) \bigg\}\sum_{p\in\Gamma_2} e^{-\pi t|p|^2/2}\, .
\label{Ggh}
\ea
Similarly, in the Coulomb phase, using Eqs.\ (\ref{ia1}-\ref{lastt}) and
\be
\I_{\T}Z_{\T}= 24\sum_{\Gamma_2}
e^{-\pi t|p|^2/2}\, ,\label{tt}
\ee
we find:
\ba
\F^I(\lambda)&=&{\lambda^2 \over
32\pi^2}\int_0^{'\infty}{dt\over t}\bigg\{32\pi^2 \sum_{\Gamma_2}
\nonumber\\
&&\hspace{-1cm}-\bigg(\sum_{a^I+a^J+\Gamma_2}s_{IJ}
+\sum_{2a^I+\Gamma_2}-16\sum_{a^I+\Gamma_2}\bigg)
\nonumber\\
&&\times{d^2 \over
d\lambda^2}\left({\pi\lambda \over \sin \pi\tilde{\lambda}}
\right)^2\bigg\} e^{-\pi t|p|^2/2}\, ,
\label{FI}
\ea
where we used the same notation as in Eq.\  (\ref{deltares}), and 
$\tilde{\lambda}\equiv \bar{p} t
\lambda/\sqrt{32G^{1/2}\mbox{Im}U}$.

It is easy to see \cite{apt2,MS2}, by following the
same steps as those described in Section 7 for the threshold corrections,
that the above type I expressions (\ref{Ggh}) and
(\ref{FI}) do appear in the Im$T\to\infty$ limit of the heterotic 
(one-loop) generating function $\F^H(\lambda, T, U, A)$, in
agreement with Eq.\ (\ref{Fgexp}). The correspondence of 
individual type I and heterotic terms is as before, with the additional
torus contribution to $\F_1^I$ [the first term on the r.h.s.\ of Eq.\
(\ref{FI})]
originating from the (P,A) orbifold sector on the heterotic side.
The leading term $4\pi\lambda^2 \mbox{Im}T$ in Eq.\ (\ref{Fgexp}) reflects
the quadratic ultraviolet divergence of $\F^H_1$.
The subleading term of order $1/{\rm Im}T$ arises also from $\F^H_1$,
and corresponds to the ``universal'' part of gravitational threshold
corrections, reproducing $\delta$ of Eqs.\ (\ref{bpstu}) and (\ref{deltares}).

\end{document}